\documentclass[aip,reprint]{revtex4-1}
\usepackage{color,graphicx,calc,url}
\usepackage{dcolumn}
\usepackage{bm}
\usepackage{amssymb}
\usepackage{amsmath}
\usepackage{wasysym}
\usepackage{mathrsfs} 
\usepackage{array}
\usepackage{mathtools}
\usepackage{xcolor}
\usepackage{commath}
\usepackage[colorlinks=true,
            linkcolor=red,
            urlcolor=blue,
            citecolor=blue]{hyperref}

\begin{document}
\title{Scaling of quadratic and linear magnetooptic Kerr effect spectra with L2$_1$ ordering of Co$_2$MnSi Heusler compound}

\author{Robin Silber}
\affiliation{Nanotechnology Centre, V\v{S}B-Technical University of Ostrava, Ostrava, Czech Republic}
\affiliation{IT4Innovations, V\v{S}B-Technical University of Ostrava, Ostrava, Czech Republic}
\affiliation{Department of Physics, Bielefeld University, Bielefeld, Germany}

\author{Daniel Kr\'al}
\affiliation{Faculty of Mathematics and Physics, Charles University, Prague, Czech Republic}

\author{Ond\v{r}ej Stejskal}
\affiliation{IT4Innovations, V\v{S}B-Technical University of Ostrava, Ostrava, Czech Republic}
\affiliation{Faculty of Mathematics and Physics, Charles University, Prague, Czech Republic}


\author{Takahide Kubota}
\affiliation{Institute for Materials Research, Tohoku University, Sendai, Japan}

\author{Yasuo Ando}
\affiliation{Department of Applied Physics, Tohoku University, Sendai, Japan}

\author{Jarom\'{i}r Pi\v{s}tora}
\affiliation{Nanotechnology Centre, V\v{S}B-Technical University of Ostrava, Ostrava, Czech Republic}
\affiliation{IT4Innovations, V\v{S}B-Technical University of Ostrava, Ostrava, Czech Republic}

\author{Martin Veis}
\affiliation{Faculty of Mathematics and Physics, Charles University, Prague, Czech Republic}

\author{Jaroslav Hamrle}
\affiliation{Faculty of Mathematics and Physics, Charles University, Prague, Czech Republic}

\author{Timo Kuschel}
\affiliation{Department of Physics, Bielefeld University, Bielefeld, Germany}



\begin{abstract}

The Heusler compound Co$_2$MnSi provides a crystallographic transition from B2 to L2$_1$ structure with increasing annealing temperature $T_a$, being a model system for investigating the influence of crystallographic ordering on structural, magnetic, optic, and magnetooptic (MO) properties. Here, we present quadratic magnetooptic Kerr effect (QMOKE) spectra depending on $\bm{M}^2$ in addition to the linear magnetooptic Kerr effect (LinMOKE) spectra being proportional to $\bm{M}$, both in the extended visible spectral range of light from 0.8\,eV to 5.5\,eV. We investigated a set of Co$_2$MnSi thin films deposited on MgO(001) substrates and annealed from 300$^\circ$C to 500$^\circ$C. The amplitude of LinMOKE and QMOKE spectra scales linearly with $T_a$, and this effect is well pronounced at the resonant peaks below 2.0\,eV of the QMOKE spectra. Furthermore, the spectra of the MO parameters, which fully describe  the MO response of Co$_2$MnSi up to the second order in $\bm{M}$, are obtained dependend on $T_a$. Finally, the spectra are compared to ab-initio calculations of a purely L2$_1$ ordered Co$_2$MnSi Heusler compound.  
 
\end{abstract}

\maketitle


Heusler compounds are a class of materials with many intriguing properties which keep them still in the focus of active research nowadays\cite{Graf2011}. The Co$_2$MnSi Heusler compound is demonstrated to be a half-metallic ferromagnet \cite{wang2005, Sakuraba2005, jourdan2014} with a band gap of 0.4 to 0.8\,eV \cite{Picozzi2002,Fuji1990} and a Curie temperature of 985\,K \cite{Brown2000}. It is well-known that the magnetic properties of Heusler compounds strongly depend on the crystallographic ordering \cite{rodan2013}. The Heusler compound Co$_2$MnSi is a good model system as it provides transition from B2 to L2$_1$ crystallographic ordering with increasing annealing temperature $T_a$ \cite{Gaier2008, Wolf2011}. The L2$_1$ structure refers to a perfectly ordered crystal, whose two Co sublattices are at 8c Wyckoff positions ($\frac{1}{4}$, $\frac{1}{4}$, $\frac{1}{4}$), whereas Mn and Si occupy the 4a (0,0,0) and 4b ($\frac{1}{2}$, $\frac{1}{2}$, $\frac{1}{2}$) positions, respectively \cite{Trudel_10a}. The B2 structure refers to lower ordering, for which interchange of atoms between Mn and Si sublattices occurs.

The magnetooptic Kerr effect (MOKE), including linear MOKE (LinMOKE) and quadratic MOKE (QMOKE), has been routinely used at single wavelengths in investigations of  magnetic properties, composition or ordering  of Heusler compounds \cite{Hamrle07b, Hamrle07a, Gaier2008, Muduli08, Muduli09, Trudel_10a, Trudel_10c, Trudel_11, Wolf2011, Veis2014, Liu2017}. In the study carried out by Wolf \textit{et al.}\cite{Wolf2011}, the dependence of LinMOKE and QMOKE on crystallographic ordering of the Co$_2$MnSi Heusler compound has been studied at single wavelength 638\,nm.  Here, we investigate LinMOKE with in-plane magnetization parallel to the plane of incidence (longitudinal MOKE, LMOKE) spectroscopy and QMOKE spectroscopy \cite{Silber2018, Silber2019} in the spectral range of 0.8--5.5\,eV. In our work, we focus on two features: (i) the QMOKE spectra show unusual oscillations for the photon energies of extended visible spectral range (ii) the LinMOKE and QMOKE scale with $T_a$.

To study this behaviour with respect to the material properties of the Co$_2$MnSi Heusler compound, the magnetooptic (MO) parameters $K$, $G_s$ and $2G_{44}$ \cite{Visnovsky1986, Silber2019} (fully describing the MO response of cubic crystal structures up to the second order in magnetization $\bm{M}$) are obtained from the LMOKE and  QMOKE spectra. The spectra of the diagonal permittivity $\varepsilon_d$ of 0th order (independent from $\bm{M}$), measured by ellipsometry, are also included. Furthermore, all the experimental results are accompanied by and compared to theoretical ab-initio calculations for L2$_1$ crystallographic ordering.


Each of the studied samples consists of a 30\,nm-thick Co$_2$MnSi layer epitaxially grown by inductively coupled plasma-assisted magnetron sputtering on a single crystalline MgO(001) substrate buffered with a 40\,nm thick Cr layer. The samples are capped with a 1.3\,nm thick Al layer. To achieve different degrees of crystallographic ordering, the samples are annealed at temperatures $T_a$ of $300^{\circ}$C, $350^{\circ}$C, $400^{\circ}$C, $450^{\circ}$C, $475^{\circ}$C and $500^{\circ}$C. The detailed structural characterisation of the samples is discussed in detail elsewhere \cite{Wolf2011} and here we just briefly summarize its results. X-ray diffraction reveals that all films are well epitaxially grown and (001) oriented. The $\theta$-$2\theta$ scan indicates B2 ordering, whereas annealing at higher temperatures promotes higher degree of L2$_1$ ordering. Namely, no L2$_1$ ordering was observed for samples annealed at 300$^\circ$C and 350$^\circ$C. A first change of crystallographic ordering towards L2$_1$ structure occurs at 400$^\circ$C and gets stronger with higher $T_a$ \cite{Gaier2008,Wolf2011}.


The analytical approximation for ferromagnetic layers describing the complex Kerr amplitude $\Phi_{s/p}$ for $s$ and $p$ polarized incident light is \cite{Hamrle07b}
\begin{subequations}
\allowdisplaybreaks
\begin{align}
\Phi_s &{}=\theta_{s}+i\epsilon_{s}= A_{s}\left(\varepsilon_{yx}-\frac{\varepsilon_{yz}\varepsilon_{zx}}{\varepsilon_{d}}\right)+B_s\varepsilon_{zx},\\[5mm]
\Phi_p &{}=\theta_{p}+i\epsilon_{p}=-A_{p}\left(\varepsilon_{xy}-\frac{\varepsilon_{zy}\varepsilon_{xz}}{\varepsilon_{d}}\right)+B_p\varepsilon_{xz},
\end{align}
\label{Kerr_analyt}%
\end{subequations}
\noindent
where $\theta_{s/p}$ is the Kerr rotation and $\epsilon_{s/p}$ is the  Kerr ellipticity.  $A_{s/p}$ and $B_{s/p}$ are the weighting optical factors, being even and odd functions of the angle of incidence (AoI), respectively. The elements $\varepsilon_{ij}$ of the permittivity tensor $\bm{\varepsilon}$ of the magnetized crystal can be described with the use of the Einstein summation as
\begin{equation}
	\varepsilon_{ij}=\varepsilon_{ij}^{(0)} \, + \, K_{ijk}M_k \, + \,\, G_{ijkl} M_k M_l\,,
\label{rozepsana permitivita}
\end{equation}
\noindent
where $M_k$, $M_l$ are the components of normalized magnetization $\bm{M}$. $\varepsilon_{ij}^{(0)}$ are the components of the permittivity tensor in 0th order in $\bm{M}$. $K_{ijk}$ and $G_{ijkl}$ are the components of the  linear and quadratic MO tensors $\bm{K}$ and $\bm{G}$, respectively \cite{Visnovsky1986}.  In the case of the cubic crystal structure 
\begin{subequations}
\begin{align}
 	&{}\varepsilon_{ij}^{(0)} = \,\delta_{ij}\varepsilon _{d}, \qquad K_{ijk} =\, \epsilon_{ijk}K,\\[1mm]
 	&{}G_{iiii}=\, G_{11}, \qquad G_{iijj}=\, G_{12}, \qquad i\neq j, \\[1mm]
 	&{}G_{1212}=\, G_{1313}=G_{2323}=G_{44},
\end{align}
\label{Einstein_tensor}%
\end{subequations} 
\noindent
with  $\delta_{ij}$ and $\epsilon_{ijk}$ being the Kronecker delta and the Levi-Civita symbol, respectively. The quadratic MO parameters $G_{11}$ and $G_{12}$ can not be separated and they contribute to MOKE signal as $G_s=G_{11}-G_{12}$. From Eqs. \eqref{Kerr_analyt}-\eqref{Einstein_tensor}, a measurement algorithm that separates LMOKE and two constituent QMOKE spectra $Q_s$ and $Q_{44}$ has been derived for (001)-oriented films and in-plane magnetization being \cite{Silber2018, Silber2019}
\begin{subequations}
\begin{alignat}{3}
	&\begin{array}{ll}
		\mathrm{LMOKE}&{=}\,\,\Phi _{s/p}^{\mu=90^{\circ}}-\Phi _{s/p}^{\mu=270^{\circ}}\\
		&{=}\,\,\pm\,2B_{s/p}K,\qquad\qquad\alpha = \mathrm{arb.\,angle}.
	\end{array}
	\label{LMOKE_sequence}
\\[3mm]	
	&\begin{array}{ll}
		 Q_s&{=}\,\,\Phi_{s/p}^{\mu=45^{\circ}}+\Phi_{s/p}^{\mu=225^{\circ}}-\Phi_{s/p}^{\mu=135^{\circ}}-\Phi_{s/p}^{\mu=315^{\circ}}\\
		 &{=}\,\,\pm\,2A_{s/p}\left(G_{s}-\frac{K^2}{\varepsilon_{d}}\right),\qquad\quad	\alpha = 45^{\circ}.
	\end{array}
	\label{QMOKE_sequence_s}
\\[3mm]
	&\begin{array}{ll}
	Q_{44}&{=}\,\,\Phi_{s/p}^{\mu=45^{\circ}}+\Phi_{s/p}^{\mu=225^{\circ}}-\Phi_{s/p}^{\mu=135^{\circ}}-\Phi_{s/p}^{\mu=315^{\circ}}\\
	&{=}\,\,\pm\,2A_{s/p}\left(2G_{44}-\frac{K^2}{\varepsilon_{d}}\right),\qquad	\alpha = 0^{\circ}.
	\end{array}
	\label{QMOKE_sequence_44}%
\end{alignat}
\end{subequations}
\noindent
The sample orientation $\alpha$ is defined as angle between [100] crystallographic direction of Co$_2$MnSi and $x$ axis of sample coordinate system. Magnetization $\bm{M}$ lies in-plane and its direction is defined by angle $\mu$, being an angle between $\bm{M}$ and x-axis of the sample coordinate system (see supplemental material I). A detailed description and derivation of the measurement algorithm can be found in Ref.\cite{Silber2019}. Finally, note that to apply the measurement algorithm according to Eqs.~\eqref{LMOKE_sequence}--\eqref{QMOKE_sequence_44}, the sample must be in full magnetic saturation for every in-plane direction of the external magnetic field. The magnetic characterization of the studied samples can be found in the supplemental material II.


\begin{figure*}
\begin{center}
\includegraphics{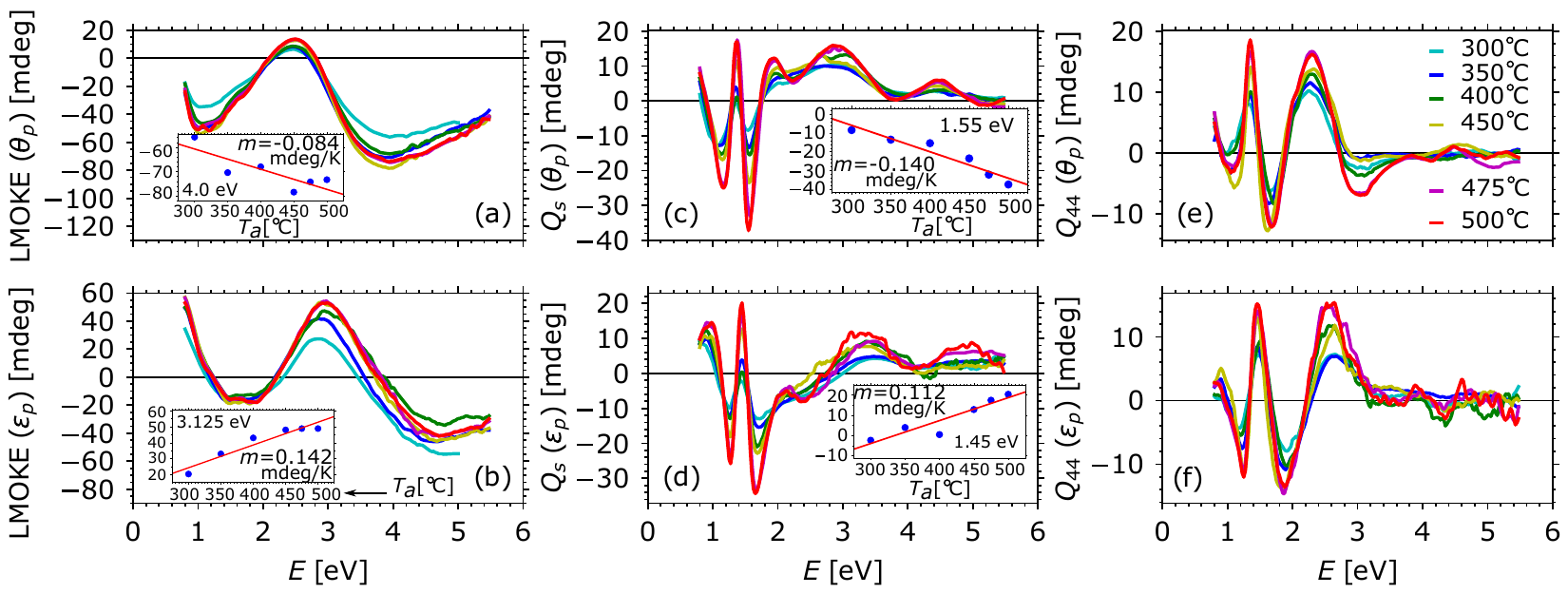}
\end{center}
\caption{LMOKE spectra of (a)  $\theta_p$, (b) $\epsilon_p$, both measured with AoI=45$^\circ$. $Q_s$ spectra of (c)  $\theta_p$ (d) $\epsilon_p$ and $Q_{44}$ spectra of (e)  $\theta_p$ and (f) $\epsilon_p$. Both measured with AoI=5$^\circ$. A $p$-polarized incident wave was used with all measurements.}
\label{fig_MO_spec}
\end{figure*}
\begin{figure}
\begin{center}
\includegraphics{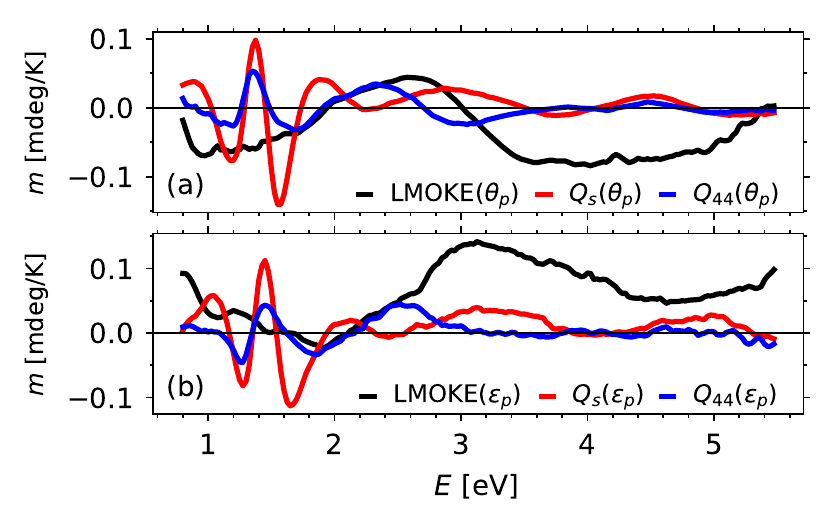}
\end{center}
\caption{Linear slope $m$ of the dependence of the linear and quadratic MOKE signal on $T_a$ plotted against photon energy for (a) $\theta_p$ and (b) $\epsilon_p$ of LMOKE, $Q_s$ and $Q_{44}$.}
\label{fig_MO_spec_slope}
\end{figure}

The LMOKE and QMOKE spectra presented in Figs.~\ref{fig_MO_spec}(a)-(f) were obtained from an in-house built MOKE spectroscopy setup \cite{Silber2018} at room temperature with $p$-polarized incident beam and using an in-plane magnetic field of 300\,mT. The spectra of LMOKE rotation $\theta_p$ and ellipticity $\epsilon_p$, measured according to Eq.~\eqref{LMOKE_sequence}, are displayed in Figs.~\ref{fig_MO_spec}(a) and (b), respectively. Here, we observe no substantial change of peak position with $T_a$, but the change of the amplitude with $T_a$ is well pronounced. In Figs.~\ref{fig_MO_spec}(c) and (d), we show $Q_s$ spectra of $\theta_p$ and $\epsilon_p$ (measured according to Eq.~\eqref{QMOKE_sequence_s}), respectively. The spectra exhibit rapid oscillations in the spectral region 0.8 - 2.0\,eV, being surprising for metallic materials. Note that the amplitude of those oscillations does depend strongly on $T_a$. Furthermore, a new peak arises in the $\theta_p$ spectra of $Q_s$ at a photon energy of 2.2\,eV and in the $\epsilon_p$ spectra of $Q_s$ at a photon energy of 2.4\,eV for higher $T_a$ values. Finally in Figs.~\ref{fig_MO_spec}(e) and (f), we show the $Q_{44}$ spectra of $\theta_p$ and $\epsilon_p$ (measured according to Eq.~\eqref{QMOKE_sequence_44}), respectively. Again, we observe a set of rapid oscillations in the spectral range 0.8 - 3.4\,eV, where the amplitude of the peaks is proportional to $T_a$. Also, the formation of a new peak with higher $T_a$ is detectable in the $\theta_p$ spectra of $Q_{44}$ at a photon energy of 4.5\,eV and in the $\epsilon_p$ spectra of $Q_{44}$ at a photon energy of 4.4\,eV. The dataset for all MOKE spectra presented in Fig.~\ref{fig_MO_spec} are accessible through Ref.\cite{data1}.


The dependence of the MOKE spectra on $T_a$ can be used as a tool for comparison of the amount of L2$_1$ ordering in Co$_2$MnSi Heusler compounds. For all MOKE spectra, we approximated the dependence on $T_a$ by linear function at each photon energy. In the peaks of QMOKE spectra, such approximation describe the experimental data very well. Figures~\ref{fig_MO_spec_slope} (a) and (b) present the value of this linear slope $m$ plotted against the photon energy for $\theta_p$ and $\epsilon_p$, respectively. In the insets of Figs.~\ref{fig_MO_spec}(a)-(d), we exemplarily present this dependence on $T_a$ together with a linear fit (red line) at photon energies with the highest slope $m$. The supplemental data files in Ref.\cite{data2} then provide set of such a graphs for the rest of photon energies. The overall highest slope $m$ can be identified for $\epsilon_p$ of LMOKE at 3.125\,eV (ca. 396\,nm), where  $m=0.142$ mdeg/K. However, $m$ possesses high values in the whole nearby region of 2.7-4.3\,eV (288-459\,nm). In case of the $\theta_p$ spectra of LMOKE, the slope of the linear dependence is well pronounced in the spectral region 3.5\,eV - 5.0\,eV (248-355\,nm) with its absolute maximum of $m=-0.084$ mdeg/K at 4.0\,eV (ca. 310\,nm). From the two presented QMOKE spectra, the $Q_s$ spectra possess a higher slope almost in the whole spectral region for both $\theta_p$ and $\epsilon_p$. Note that at 1.55\,eV (800\,nm) the $\theta_p$ spectra of $Q_s$ possess a slope of $m=-0.140$ mdeg/K, being nearly equal (in absolute quantities) to the overall highest slope value of all MOKE spectra.  Although $Q_s$ requires the measurement of Kerr angles under four different $\bm{M}$ directions, whereas LMOKE requires only two (compare Eqs.~\eqref{LMOKE_sequence} and \eqref{QMOKE_sequence_s}), there are widely available laser diodes at 785\,nm (ca. 1.53\,eV) and 808\,nm (ca. 1.57\,eV). 
 
To obtain spectra of MO parameters $K$, $G_s$ and $2G_{44}$ to the permittivity tensor of the Co$_2$MnSi layer, the experimental spectra were described numerically using 4x4 Yeh matrix formalism, in which coherent propagation of the electromagnetic plane waves through a multilayer system is considered \cite{Yeh1980}. The MO parameters of the Co$_2$MnSi layers were treated as free parameters of the model and $\varepsilon_d$ was obtained from the ellipsometry measurements (supplemental material III). 
\begin{figure*}
\begin{center}
\includegraphics{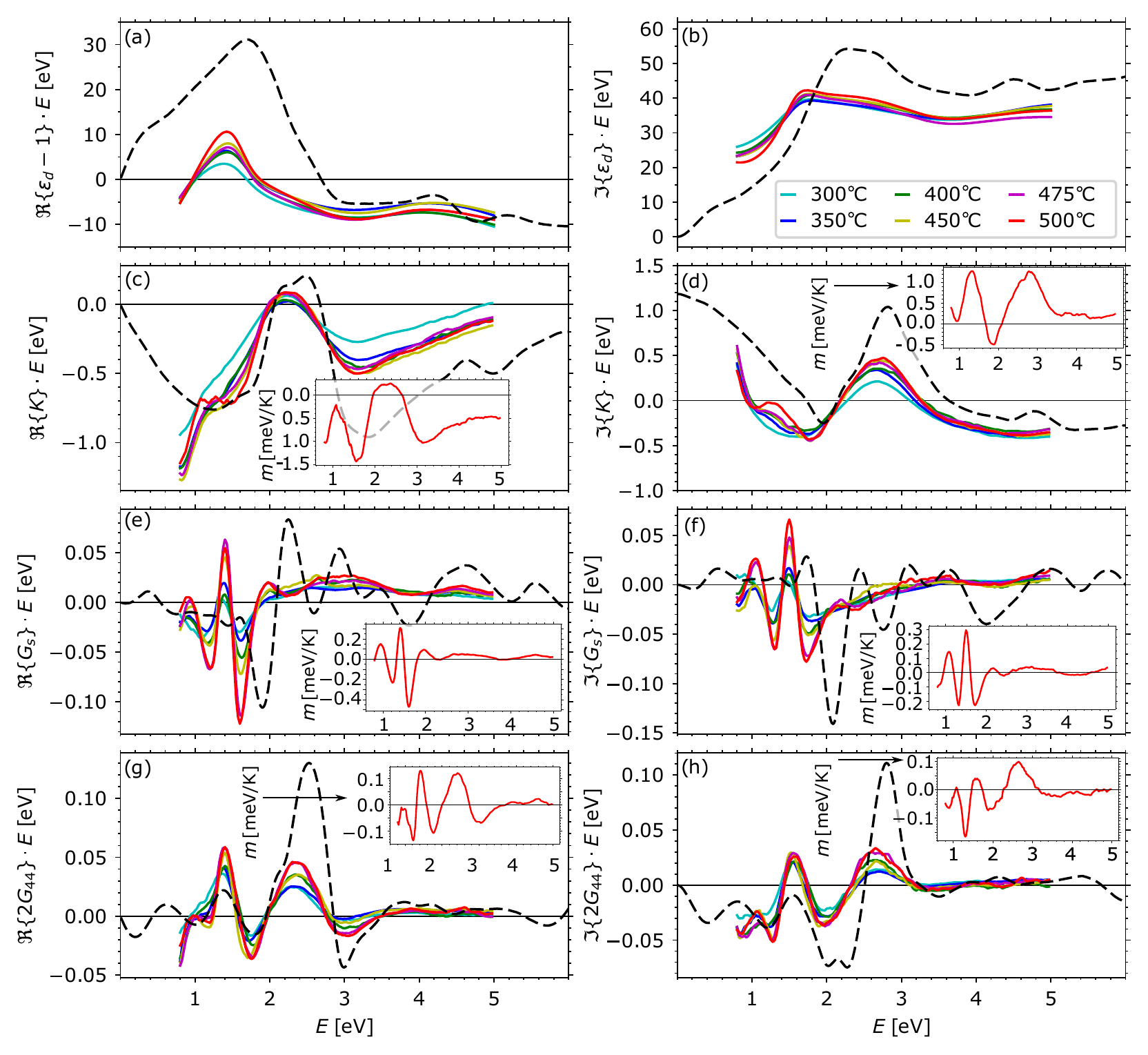}
\end{center}
\caption{Spectra of (a) real and (b) imaginary part of $\varepsilon_d$ (permittivity of 0th order in $\bm{M}$), (c) real and (d) imaginary part of linear MO parameter $K$, (e) real and (f) imaginary part of quadratic MO parameter $G_s$, (g) real and (h) imaginary part of quadratic MO parameter $2G_{44}$. Coloured full lines are the experimental data and black dashed lines are ab-initio.}
\label{fig_MOpar}
\end{figure*}

The resulted spectra of $\varepsilon_d$ (permittivity in the 0th order in $\bm{M}$) and of the MO parameters $K$, $G_s$ and $2G_{44}$ are shown in Fig.~\ref{fig_MOpar} and are accessible as supplemental data file in Ref.\cite{data3}. All experimental spectra are accompanied by theoretical ab-initio calculations (dashed lines). For details about the ab-initio calculations, please see supplemental material IV including Refs. \cite{Wien2k, Kubo57, Draxl2006, Hamrlova2016}. Note that all spectra shown in Fig.~\ref{fig_MOpar}  are in the form multiplied by photon energy $E=\hbar\omega$, being actually proportional to the conductivity spectra $\sigma_{ij}$, since $\varepsilon_{ij}=\delta_{ij}+i\sigma_{ij}\hbar/(\varepsilon_0 E)$.

 The real and imaginary part of the complex spectra of $\varepsilon_d$ are shown in Figs.~\ref{fig_MOpar} (a) and (b), respectively. The theoretical spectra of $\varepsilon_d$ predict the major optical transitions to appear for higher energies than experimentally observed, which is apparent from Fig.~\ref{fig_MOpar}(b). This discrepancy can be arguably attributed to the employed exchange-correlation potential, the major approximation present in the density functional theory (DFT), which still slightly overestimates distance between occupied and excited electronic states by about 0.3\,eV \cite{Cohen2008,Cramer2009}.

The complex spectra of the linear MO parameter $K$ are shown in Figs. \ref{fig_MOpar} (c) and (d). Here, a new peak emerges with the presence of L2$_1$ ordering in the spectral region of 1.5\,eV and 1.3\,eV in real and imaginary part, respectively (note that this peak was not well observable in the LMOKE spectra itself).  Although this peak is not very pronounced in the ab-initio calculations, the experimental and theoretical spectra provide otherwise good agreement.

 In Figs.~\ref{fig_MOpar} (e) - (h), we show the complex spectra of $G_s$ and $2G_{44}$. When compared to the spectra of $Q_s$ and $Q_{44}$, described above, no new features are observed here. Although theoretical calculations of $G_s$ are shifted in photon energies, similarly to $\varepsilon_d$ spectra,  amplitude and period of those oscillations are well described ab-initio. In the case of $2G_{44}$ spectra, the theory is in reasonable match with the experiment. Note that theoretical spectra of $K$ and $2G_{44}$, for which the ab-initio calculations describe the experimental spectra well, are calculated from off-diagonal elements $\varepsilon_{ij} (i\neq j$) of the permittivity tensor, whereas spectra of $G_s$ are calculated as small variation of diagonal permittivity $\varepsilon_{ii}$ (see supplemental material IV).

In the insets of Figs.~\ref{fig_MOpar} (c)-(h) we present the spectra of slope $m$ of the linear dependence of MO parameters (multiplied by photon energy $E$) on $T_a$. The highest values of $m$ can be found for the MO parameter $K$, although when values of $m$ are compared to the values of its MO parameter, it is the slope $m$ of $G_s$ that is most pronounced. 


In conclusion, we investigated QMOKE and LMOKE spectroscopy of Co$_2$MnSi which undergoes a transition from B2 to L2$_1$ ordering provided by increasing annealing temperature $T_a$. We showed that the amplitude of the spectra depends on $T_a$ and this dependence can be reasonably approximated by linear function, especially in the peaks of QMOKE spectra. We suggest the photon energies where this dependence is most pronounced and can be used as an indirect sensor of the structure ordering from an application point of view. Apart from the increase of the amplitude, several new peaks emerged with increasing L2$_1$ ordering. We further extract the spectra of the MO parameters $K$, $G_s$ and $2G_{44}$ and describe them ab-initio. In the case of $K$ and $2G_{44}$ spectra, the ab-initio calculations describe the experimental spectra well. The theoretical spectra of $\varepsilon_d$ and $G_s$ are considerably shifted in photon energy with respect to the experimental spectra, which we attributed to the used exchange-correlation potential. 

This work was supported by projects:   RE1052/42-1(DFG), GA18-22102S (GACR), CZ. $02.1.01/0.0/0.0/16\_013/0001791$(EU),  CZ.$02.1.01/0.0/0.0/15\_003/0000487$ (EU) and by EHP-CZ-ICP-1-013 (EEA grants).



\renewcommand{\section}[2]{}%
\bibliography{main.bib}

\end{document}